\documentclass[%
 reprint,
 amsmath,amssymb,
 aps,
]{revtex4-2}

\usepackage{ulem}
\usepackage{graphicx}
\usepackage{dcolumn}
\usepackage{bm}
\usepackage{graphicx} 
\usepackage{physics}
\usepackage{array}
\usepackage{amsmath}
\usepackage{siunitx}

\usepackage{hyperref}

\DeclareSIUnit\atom{atom}

\usepackage[mathlines]{lineno}

\DeclareSIUnit\gauss{G}
\newcommand{\dyads}[3]{\ket{#1}_{#3}\!\bra{#2}}

\begin{document}


\title{Isolating Pure Quadratic Zeeman Splitting}

\newcommand{\JUAddress}{Institute of Physics, Jagiellonian University in
Krak\'ow, \L{}ojasiewicza 11, 30-348 Krak\'ow, Poland}

\newcommand{\JUDSAddress}{Jagiellonian University, Doctoral School of Exact and Natural Sciences, Faculty of Physics, Astronomy and Applied Computer Sciences, \L{}ojasiewicza 11, 30-348 Krak\'ow, Poland}

\author{Arash Dezhang Fard}
\email{arash.dezhangfard@doctoral.uj.edu.pl}
\affiliation{\JUAddress}
\affiliation{\JUDSAddress}
 
\author{Marek Kopciuch}%
\email{marek.kopciuch@doctoral.uj.edu.pl}
\affiliation{\JUAddress}

\author{Yujie Sun}
\affiliation{\JUAddress}

\author{Przemys\l aw W\l{}odarczyk}
\affiliation{\JUAddress}

\author{Szymon Pustelny}
\affiliation{\JUAddress}

\date{\today}

\begin{abstract}
Nonlinear magnetic interactions provide access to complex quantum spin dynamics and thus enable the study of intriguing physical phenomena. However, these interactions are often dominated by the linear Zeeman effect, which can complicate system dynamics and make their analysis more challenging. In this article, we theoretically and experimentally introduce a method to induce the quadratic Zeeman effect while effectively compensating for its linear counterpart. By isolating the quadratic Zeeman contributions, we demonstrate and analyze controlled superposition generation between specific magnetic sublevels in room-temperature rubidium-87 atoms. This study opens avenues for controlling any spin system, regardless of its total angular momentum, which we plan to explore further in the context of quantum-state tomography and engineering (e.g., spin squeezing).
\end{abstract}

\maketitle

\section{Introduction}

At the core of many quantum algorithms lies the ability to selectively address specific transitions within a target system to modify quantum information \cite{Nielsen2010, Adam2015}. This capability, however, presents challenges across various platforms, including collective atomic ensembles like warm atomic vapors, where more than two states are used for information storage (qudits) \cite{Chaudhury2007Quantum, Colangelo2013, marek}. Typically, lifting the degeneracy of magnetic transitions between Zeeman sublevels involves applying strong magnetic fields to induce the nonlinear Zeeman effect. This approach, however, mainly results in a strong linear Zeeman effect, where the degeneracy is still present, which can be detrimental to fidelity of the performed operations. It may also lead to magnetic-field inhomogeneity, resulting in enhanced relaxation/infidelity, and is incompatible with magnetic shielding, often used to provide better control over the magnetic-field environment \cite{Budker2013Optical}. An alternative method for phase imprinting on specific transitions utilizes sequences of weak magnetic-field pulses combined with nonlinear evolution induced by the AC Stark \cite{Chaudhury2007Quantum, Echaniz2007HamiltonianDI, Pawela2012QuantumCW}. Here, both the technique infidelity and complexity may be an important issue. In our work, we propose a method to achieve this nonlinearity by leveraging the quadratic Zeeman effect while compensating for the linear Zeeman effect.

In 1896, Pieter Zeeman discovered that spectral lines emitted from sodium vapor split when subjected to a magnetic field \cite{1897ApJ.....5..332Z}. This phenomenon, known today as the Zeeman effect, arises from the magnetic-field-induced splitting of energy sublevels. In the ground state of alkali-metal atoms, where the total angular momentum of the electron is $1/2$, the magnetic-field dependence of the energy $E_{\ket{F,m_F}}$ of a given Zeeman sublevel $m_F$ of the hyperfine state of the total angular momentum $F$ can be analytically derived. It results in the well-known Breit-Rabi formula \cite{BreitRabi}
\begin{equation}
    \begin{split}
        E_{\ket{F,m_F}} = - \frac{\Delta E_\text{hfs}}{2(2I+1)} +& g_I \mu_{B} m_F B
        \\ \pm& \frac{\Delta E_\text{hfs}}{2} \sqrt{1+\frac{4m_Fx}{2I+1}+x^2}\,,
    \label{eq:Breit_Rabi}
    \end{split}
\end{equation}
where $\Delta E_\text{hfs}$ is the hyperfine splitting, $I$ is the nuclear spin, $g_I$ is the nuclear $g$-factor, $\mu_B$ is the Bohr magneton, and $B$ is the magnetic field. The parameter $x=\frac{(g_J+g_I)\mu_B B}{\Delta E_\text{hfs}}$, with $g_J$ being the Land\'e factor for the electron, determines the response of the system to the external magnetic field $B$ and $\pm$ is given by the hyperfine level $F=I\pm 1/2$ that the atom resides in.  

The Zeeman effect is often analyzed within the linear regime, where the magnetic field is sufficiently weak for the Breit-Rabi formula to be expanded only up to the lowest (linear) order of the magnetic field. This regime is the regime where most optically-pumped magnetometers operate \cite{budker2007optical,Budker2013Optical,Budker2000Sensitive,allred2002high,alexandrov2003recent,10.1063/1.1839274,Groeger2006High,PhysRevA.75.051407,pustelny2008magnetometry,kim2016ultra,colombo2016four,oelsner2019performance,bevington2019imaging,PhysRevA.101.053427,fang2020high}. These devices, being the most sensitive magnetic-fields sensors \cite{chalupczak2012room,kominis2003subfemtotesla}, measure optical signals that linearly depend on the magnetic field. However, as the field strength increases, higher-order terms of the Breit-Rabi-formula expansion must be considered. In this case, the nonlinear contribution can no longer be neglected, which results not only in the nonlinear dependence of the optical signal on the magnetic field \cite{PhysRevA.73.053404,PhysRevA.82.023417}, but often in the deterioration of the signal \cite{gawlik2017nonlinear} and the generation of such systematics as heading errors \cite{headingerror} in the magnetometers.

Despite their detrimental role in magnetometry, nonlinear magnetic perturbations give rise to a variety of interesting phenomena. Two representative examples are orientation-to-alignment conversion (OAC) \cite{rochester2001atomic} and alignment-to-orientation conversion (AOC) \cite{PhysRevLett.85.2088,PhysRevA.102.053102,PhysRevLett.97.043002}, which are closely related to the ability to induce spin squeezing in atoms \cite{PhysRevA.85.022125}. In this context, the nonlinear Zeeman interaction can be considered a valuable resource in quantum metrology and engineering. It should be noted, however, that an obstacle to the application of these nonlinear terms is the presence of linear effects, which may dominate and undermine their usefulness.

In this paper, we present a novel technique that enables the controlled evolution of atomic polarization due to the quadratic Zeeman effect while simultaneously canceling the contribution from the linear Zeeman effect. This is achieved by applying a pulse of an oscillating magnetic field to the atomic system, ensuring that the pulse contains a large integer number of oscillations. Our approach, based on an $RLC$ circuit, is tested using nonlinear magneto-optical rotation (NMOR) \cite{RevModPhys.74.1153} observed in room-temperature rubidium-87 vapor. By studying the evolution of NMOR signals as a function of pulse amplitude and duration, we demonstrate the technique’s ability to imprint an arbitrary phase between magnetic sublevels. This control over the quantum state of atoms, along with the capability to generate arbitrary states in the long-lived ground level, makes the technique promising for applications in quantum-state engineering and quantum metrology.

\section{Magnetic-field pulser}

\begin{figure}
    \centering
    \includegraphics[width=\columnwidth]{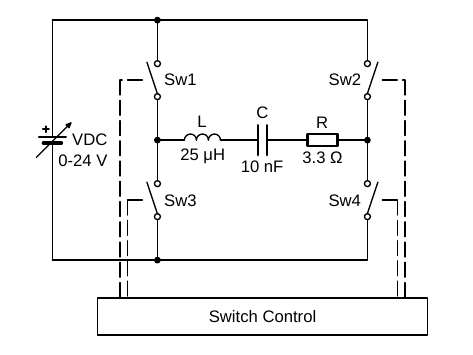}
    \caption{Schematic diagram of the pulse generator. Sw1 to Sw4 are the switches forming the H-bridge that drives the $RLC$ circuit. The circuit consists of the Helmholtz coils with inductance $L$ used to generate the magnetic field, a capacitor $C$ that allows for tuning of the resonant frequency, and a resistance $R$ (originating from the coils and an additional resistor) that allows for control over the maximum current of the system (but also affecting its resonance frequency). During the pulse, the switch control alternates between two sets of positions of the switches Sw1-Sw4 and Sw2-Sw3, pumping energy into the RLC circuit in the form of AC current. This current is supplied by the variable DC power supply. Prior to and after the pulse Sw3 and Sw4 switches are kept closed, so that no energy is pumped into the circuit.}\label{fig:circuit}
\end{figure}

The schematic diagram of the electronic circuit used to generate pulses of oscillating magnetic field is shown in Fig.~\ref{fig:circuit}.
The pulser is essentially a resonant $RLC$ circuit, where the inductance $L$ is determined by the coils used to generate the magnetic field experienced by the atoms undergoing the Zeeman effect. This inductance is connected in series with a resistor with resistance $R$ and a capacitor with capacitance $C$. The resistor limits the current that flows through the coils and adjusts the quality factor of the $RLC$ circuit. Simultaneously, the capacitance $C$ determines the system resonance frequency.

The $RLC$ circuit is powered by four switches (Sw1 - Sw4) configured in an H-bridge arrangement. In this design, the switches are realized with field-effect power transistors governed by a switch control system. This system, being an Arduino-based generator produces a sequence of square-like pulses with adjustable frequency and duration. By alternating the signal between positive and negative voltages, the transistors are toggled in pairs (Sw1 and Sw4, Sw2 and Sw3), producing the alternating current that drives the $RLC$ circuit. Prior to activation, Sw3 and Sw4 remain closed, ensuring that no current flows through the coil. Upon triggering, the switches flip synchronously, charging the coils and generating alternating current. After the pulse ends, the lower transistors (Sw3 and Sw4) close again, and the system dissipates the stored energy as a damped AC current. This design ensures that the total current that flows through the coils during the entire pulse integrates to zero.

After the system is triggering and the first pair of switches closes at $t=0$, the circuit starts oscillating, storing energy. Under our conditions, an oscillation frequency of $\SI{326}{\kilo\hertz}$, predetermined by the Arduino-based generator, begins to build up with a characteristic rate $\Gamma = \SI{2e5}{\per\second}$. However, a steady increase in the amplitude is observed throughout the duration of the pulse. At time $t = \tau$, where $\tau$ indicates the length of the pulse, that in the considered case is $\SI{100}{\micro\second}$, the switches open and the current in the circuit begins to decay at the rate $\Gamma$ while oscillating with the same frequency.

\begin{figure}[h!]
    \centering
    \includegraphics[width=\columnwidth]{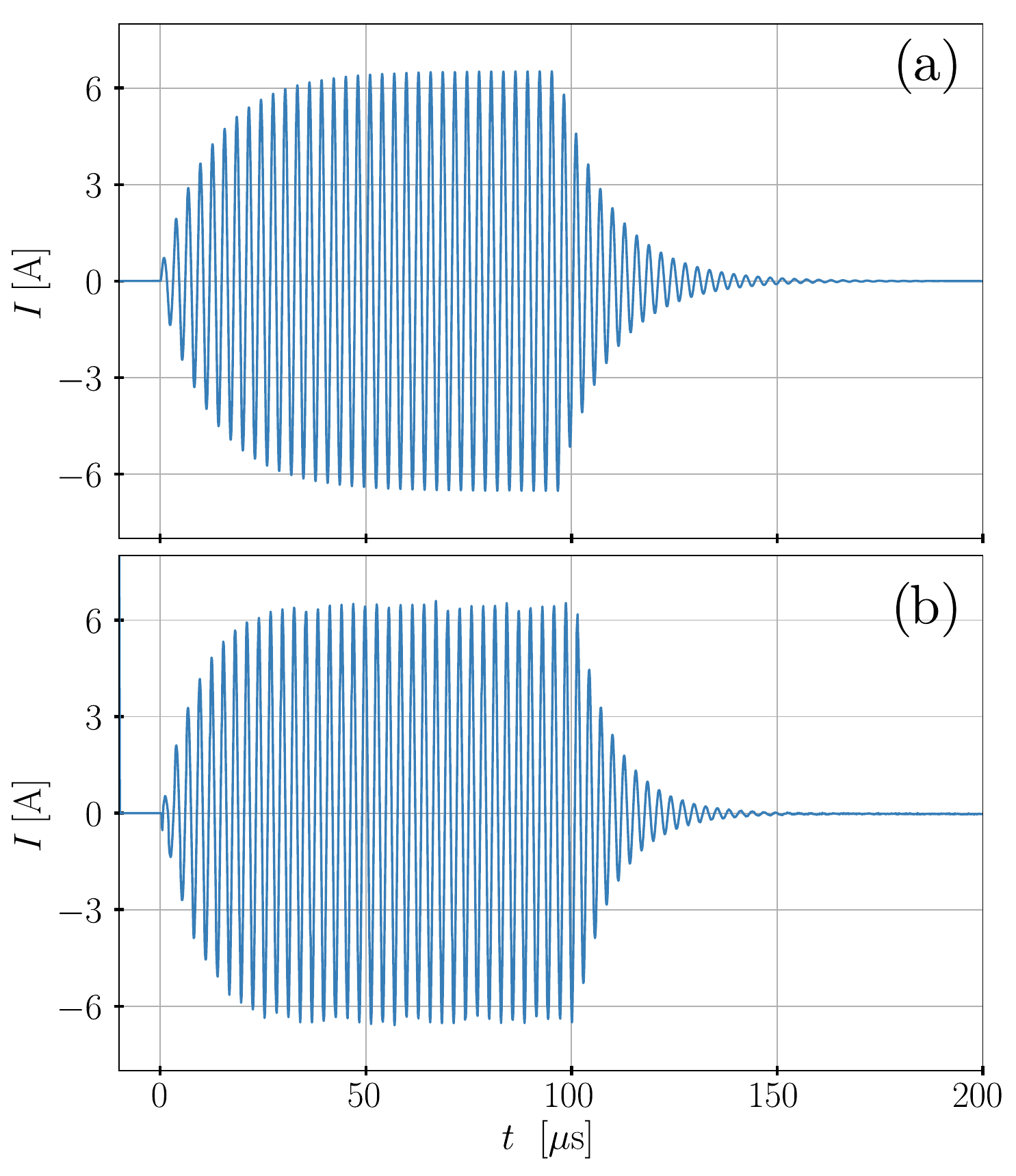}
    \caption{Simulated (a) and measured (b) currents through the coils. Both signals exhibit a similar character; starting at $t=0$, they increase to a maximum current amplitude of approximately $\SI{7.0}{\ampere}$ (slightly less in the case of simulations) within $\approx\SI{20}{\micro\second}$. After about $\SI{100}{\micro\second}$, the bridge is switched off and the system irradiates energy as damped AC current. The simulations/measurements were performed with following parameters: $R = \SI{3.3}\ohm$, $C = \SI{10}{\nano\farad}$, $L= \SI{25}{\micro\henry}$, and $V=\SI{23}{\volt}$.}
    \label{fig:MF_osci}
\end{figure}

Figure~\ref{fig:MF_osci}(a) shows a simulation of the current over the coil generated by the circuit presented in Fig.~\ref{fig:circuit}. The simulation was performed using SPICE-based software with an input voltage $V = \SI{23}{\volt}$, a capacitance $C=\SI{10}{\nano\farad}$, a inductance $L = \SI{25}{\micro\henry}$, and resistance $R = \SI{3.3}{\ohm}$, with the current is measured across the $RLC$ circuit. At the same time, the current measured in a real circuit, incorporating the same element values, is presented in Fig.~\ref{fig:MF_osci}(b). The measured trace is similar to the simulated one. Specifically, we observe the buildup of a 326-kHz AC current (frequency determined by the Arduino generator). After the upper switches close, there is a successive leveling up of the oscillation amplitude at roughly $\SI{6.5}{\ampere}$, also predicted by the simulation (please note that here the H-bridge was supplied with a voltage of $\SI{23}{\volt}$). Finally, a decrease in the current is observed when the switches are opened again.

\section{Cancellation of dynamics induced by the linear Zeeman effect}

To consider the magnetic-field induced evolution of the system, let us expand the Breit-Rabi formula [Eq.~\eqref{eq:Breit_Rabi}] up to the quadratic term in the magnetic field $B$. Neglecting the scalar term, which shifts all the levels of a particular hyperfine state in the same manner, the energy-level shift is given by
\begin{eqnarray}
    \Delta E_{\ket{F,m_F}}(t)&=&\pm \left[\frac{g_J\mu_B}{2I+1} m_{F}B(t) \right.\nonumber\\
    && \quad -\left.\frac{g_J^{2}\mu_{B}^{2}}{\Delta E_{hfs}(2I+1)^2}m_F^2 B^{2}(t)\right]\\
    &=&\pm\hbar \left[ \Omega_L^{(1)}(t) m_{F} - \Omega_L^{(2)}(t) m_{F}^{2}\right]\nonumber
    \label{eq:LevelSplitting}
\end{eqnarray}
where $\pm$ corresponds to the hyperfine state $F=I\pm1/2$ and $\Omega_L^{(1)}$ and $\Omega_L^{(2)}$ are the Larmor-frequency contributions due to the linear and quadratic Zeeman effect, respectively. Note that we neglected the nuclear contribution to the Zeeman splitting because $g_I\ll g_J$.

\begin{figure}[h!]
    \centering
    \includegraphics[width=\columnwidth]{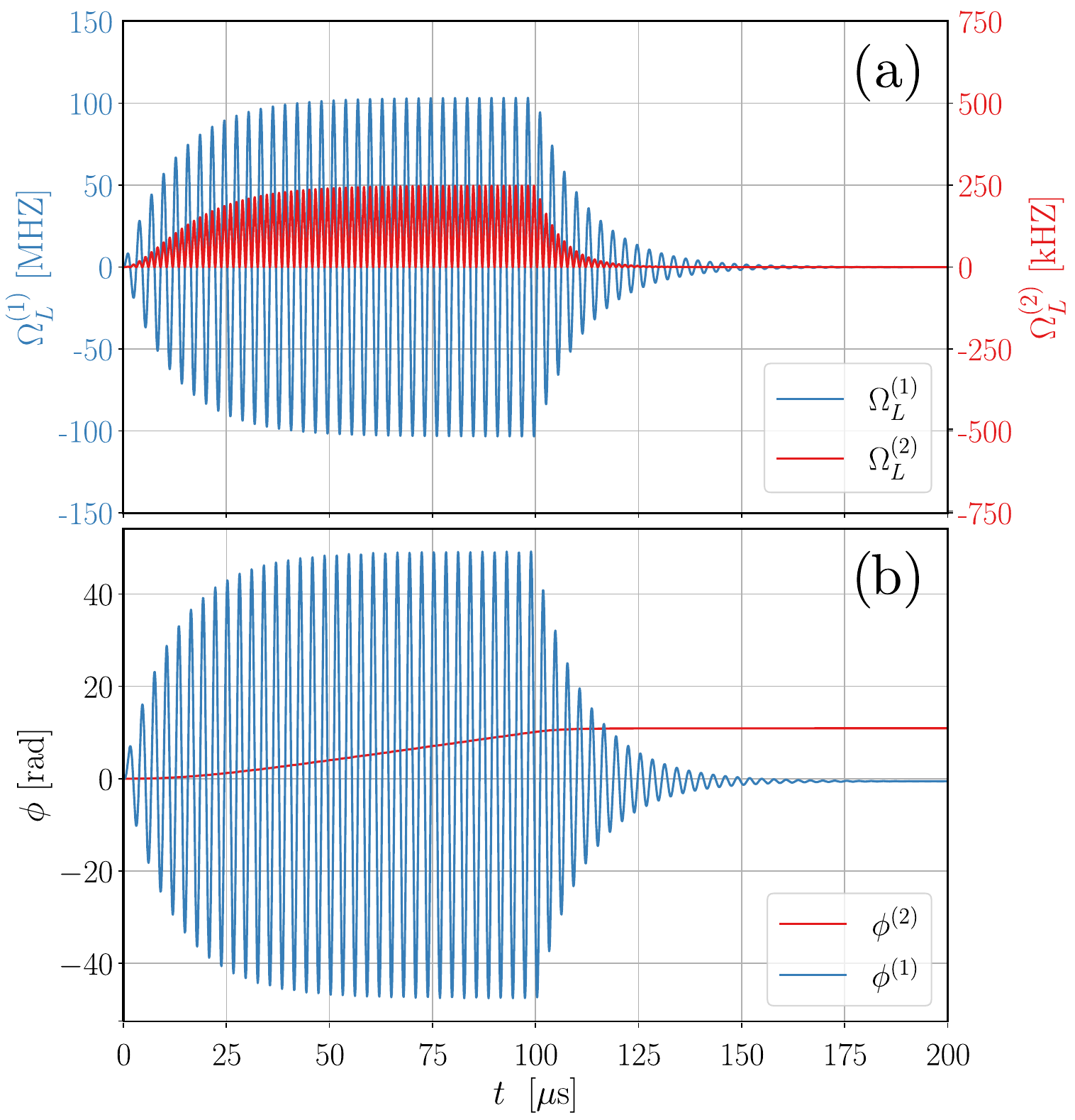}
    \caption{Simulated contributions to the Larmor frequency due to the linear (blue) and quadratic (red) Zeeman effects (a) and the corresponding phases accumulated by the atoms (b). Despite the large difference in the precession frequency (note different scales), the accumulated phase due to the linear effect averages out, while the quadratic contribution builds up. The simulations were performed for the pulse with the parameters presented in Fig.~\ref{fig:MF_osci}.}
    \label{fig:Zeeman_effect}
\end{figure}

The phase gained by a precessing spin during the magnetic-field pulse can be written as 
\begin{equation}
    \phi^{(k)}(\tau) = \int_{0}^{\infty} \Omega_L^{(k)}(t)\dd{t},
\label{eq:phase_def}
\end{equation}
where $k=1,2$ determines the linear and quadratic contribution, respectively. Note that the magnetic field $B(t)$ is proportional to the current $I(t)$ and hence the precession, dominated by the linear effect, changes sign every half of the oscillation cycle of the current. If the total (time-integrated) current through the coils is zero, the linear contribution disappears ($\phi^{(1)}=0$). Simultaneously, the quadratic contribution, that is proportional to the square of the magnetic field, does not average out ($\phi^{(2)}\neq 0$). This is illustrated in Fig.~\ref{fig:Zeeman_effect}(a), where the linear and quadratic contributions to the Larmor frequency are presented for the pulse of the current (Fig.~\ref{fig:MF_osci}) and the described geometry. Although the amplitude of the linear contribution is nearly three orders of magnitude larger than its quadratic counterpart, its sign oscillates between positive and negative values. At the same time, the sign is always positive for the quadratic contribution. In turn, the accumulated phase of the quadratic contribution is larger than that of the linear contribution [Fig.~\ref{fig:Zeeman_effect}(b)].

Due to the significantly different contributions to the Larmor frequency from the linear and quadratic Zeeman effects, an important question is: To what extent the linear contribution is compensated? In this context, it is important to note that the design of the pulser, utilizing the $RLC$ circuit, ensures that the total current through the coils is zero. This can be demonstrated by observing that both prior to and after the pulse, the charge on the capacitor is zero. Since charge is the integral of the current over time, this confirms that the total current through the coils is zero, and thus, the linear phase contribution $\phi^{(1)}$ is perfectly compensated.

Finally, it is important to stress that the above considerations are valid for motionless atoms. In the case of a room-temperature vapor, however, atoms are freely moving across the cell, bouncing off its walls. In this scenario, they may experience gradients of the pulse-induced magnetic field, which could affect their evolution, leading to excessive relaxation of atomic polarization. Below, this effect is discussed in more detail.

\section{Experimental implementation}

The experiment is designed to test our cancellation method using nonlinear magneto-optical rotation, which is characterized by the light-intensity-dependent polarization rotation of linearly polarized light as it propagates through a medium subjected to a longitudinal magnetic field. In our approach, the process is divided into three distinct stages. In the first stage, an initial quantum state is prepared using optical pumping. During the following stage, this state is modified using our magnetic-field pulser. Finally, in the last stage, the polarization rotation of a weak (non-perturbing) light is measured when atoms are placed in a weak magnetic field.

As shown in Refs.~\cite{marek, PhysRevA.109.032402, Piotrak2024perfectquantum}, when probing with linearly polarized light, propagating along the (probing) magnetic field, the polarization-rotation signal contains three components: a non-oscillating component and two components oscillating at twice the Larmor frequency (all decaying over time). In the $F=1$ state, the signal at twice the Larmor frequency is determined by a coherence between the $\ket{1,\pm 1}_{z}$ Zeeman sublevels
\begin{equation}
    \hat{\alpha}_R \propto \dyads{1,1}{1,-1}{z} + \dyads{1,-1}{1,1}{z},
    \label{eq:observable}
\end{equation}
where the subscript $z$ (and $x$ and $y$ later on) indicates the direction of the quantization axis in which the state is described. Observation of the amplitude and phase of the coherence provides then insight into the coherence.

\begin{figure*}
    \centering
    \includegraphics[width=1.95\columnwidth]{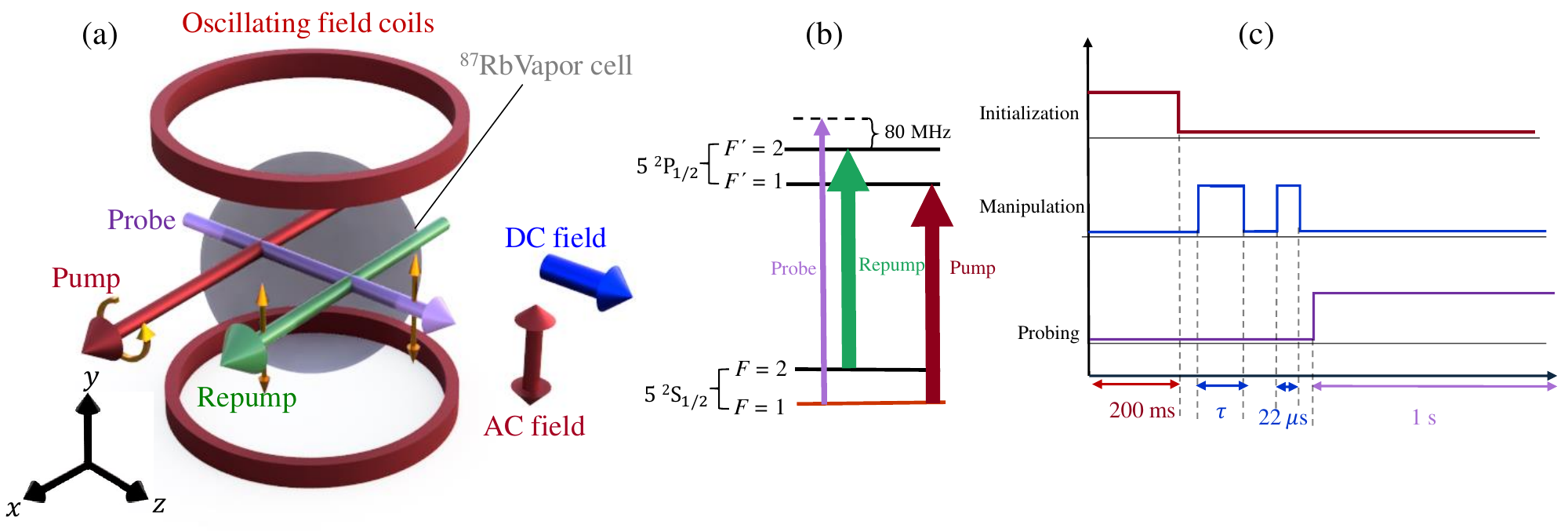}
    \caption{(a) Schematic of the experimental setup. The green, red, and purple single-headed arrows illustrate the propagation directions of the repump, pump, and probe beams, respectively. The smaller orange arrows show the polarizations of the beams. The blue arrow illustrates the direction of the static magnetic field present during probing, which coincides with the quantization axis at the stage. The double-headed red arrow illustrates the oscillating magnetic field. (b) Energy-level diagram associated with the rubidium-87 D$_1$ line, with marked laser tunings. (c) Experimental sequence used in our measurements. The initial state is created by the application of a circularly polarized pump laser simultaneously with a linearly polarized repump laser. The state is then manipulated by the application of an oscillating magnetic field for a given time $\tau$. Finally, the state is measured using the probe laser.}
    \label{fig:ExperimentalSetup}
\end{figure*}

The schematic of the experimental setup used to test the cancellation method is shown in Fig.~\ref{fig:ExperimentalSetup}.
The heart of the system is a paraffin-coated spherical atomic vapor cell with a diameter of $\SI{37}{\milli\metre}$. The cell contains atomic vapor of $^{87}$Rb and is kept at a temperature of approximately $\SI{38}{\celsius}$, corresponding to a concentration of $9\times 10^{9}$~atoms/cm$^3$. The cell is housed inside a multilayer magnetic shield composed of three nested cylindrical layers of mumetal and an innermost cubic layer of ferrite, providing a magnetic-field shielding factor of about $10^{6}$. This configuration enables us to maintain stable magnetic conditions within an innermost volume of $\SI{19.4}{\centi\metre}\times \SI{19.4}{\centi\metre}\times \SI{19.4}{\centi\metre}$. Three solenoids and three sets of anti-Helmhotz coils, oriented along $x$, $y$, and $z$, are mounted inside the shield to compensate for any residual magnetic fields (homogeneous and linear gradients), as well as to produce weak DC magnetic-field pulses in the linear Zeeman regime (note that these are different coils than the ones used for the magnetic-field pulser).

The crucial component of the setup is an additional set of Helmholtz coils, which are used to generate an oscillating magnetic field with an amplitude exceeding $\SI{25}{\gauss}$.
This set consists of two circular coils, each $\SI{50}{\milli\metre}$ in diameter and having 10 turns. The coils are wound using a wire that is \SI{2}{\milli\metre} in diameter, resulting in the coils having a resistance of $\SI{0.1}{\ohm}$ and an inductance of $\SI{25}{\micro\henry}$.

State preparation and probing are realized using three lasers (Fig.~\ref{fig:ExperimentalSetup}). The first laser (the pump), tuned to the $F=1\rightarrow F'=1$ transition is utilized to generate the initial state in the atoms. The second laser (the repump), tuned to the $F=2\rightarrow F'=2$ transition, is used to deplete the $F=2$ manifold, ensuring the $F=1$ state is fully occupied. Finally, the third laser (the probe), blue-detuned from the center of the Doppler broadened $F = 1\rightarrow F' = 2$ transition by $\SI{80}{\mega\hertz}$, is employed to measure polarization rotation in the atoms. In this way, it detects the phase imprinted onto the atoms with our method. The experimental sequence is as follows: first, interaction with the pump laser creates the initial state. This process is conducted with the repump laser on. Next, an oscillating magnetic field is applied using our pulse generator to manipulate the quantum state of the atoms (see more details below). Finally, the state measurement is performed based on the polarization rotation of the probe laser (for more details regarding the experimental setup and procedure, see Ref.~\cite{PhysRevA.109.032402}).

\section{Results and discussion}

\begin{figure}
    \centering
    \includegraphics[width=0.9\columnwidth]{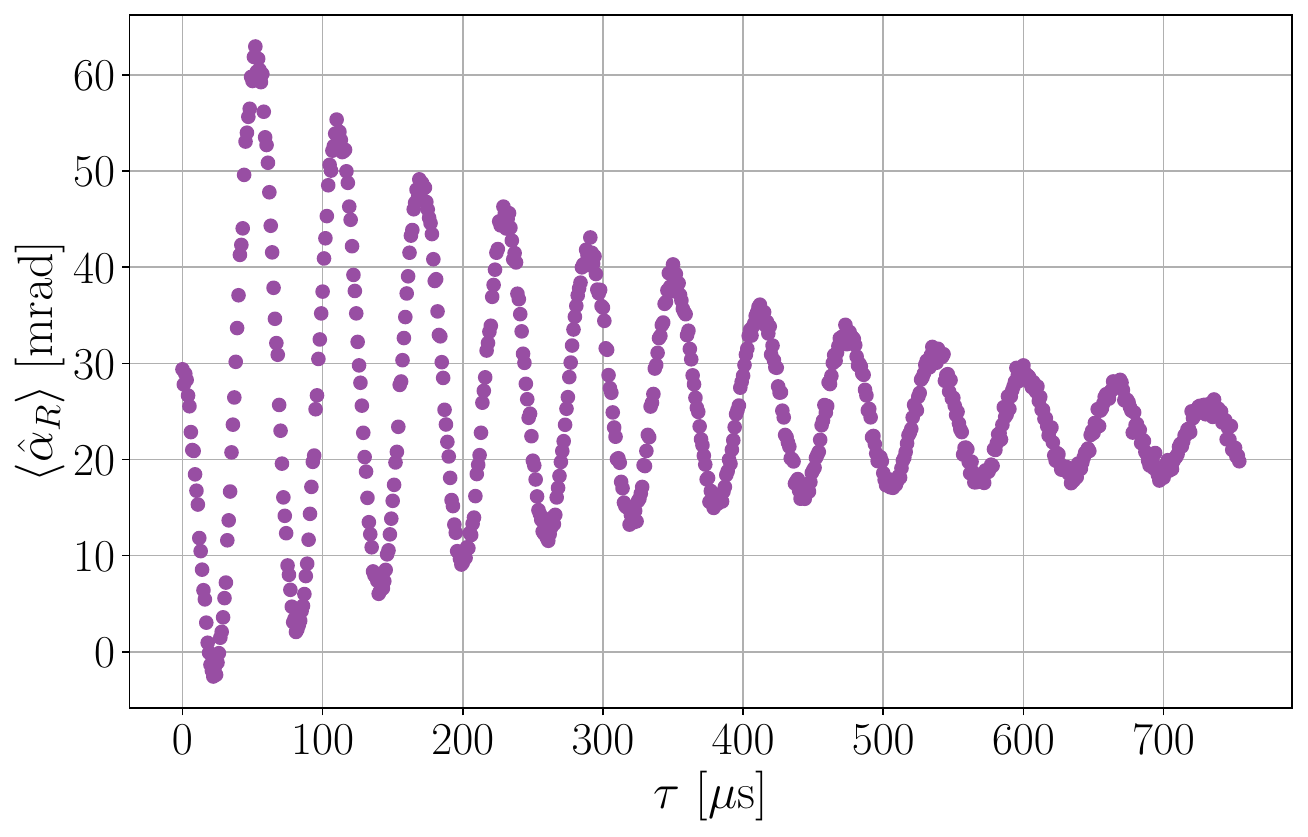}
    \caption{The value of the measured observable $\expval{\hat{\alpha}_R}$ as a function of the pulse duration $\tau$. With prolongation of the pulse, the amplitude exhibits oscillations, which period is determined by the amplitude of the pulse, in presented case the driving amplitude was set to $\SI{24}{\volt}$. In addition to the oscillations of the magnetic field, the amplitude deteriorates, which stems from the inhomogeneities of the magnetic field induced with the pulser.}
    \label{fig:alphaR_vs_tau}
\end{figure}

To validate our technique, we choose the $\ket{1,1}_{x}$ state, which can be generated using circularly polarized pump light, propagating along the $x$-axis and tuned to the $F=1\rightarrow F'=1$ transition. It can be shown that by applying the oscillating magnetic field pulse to the state, and successively rotating the state with the ``conventional'' pulse by $\pi/4$ around the $x$-axis, one can obtain the polarization rotation signal depending explicitly on the quadratic phase $\phi^{(2)}$ (see Appendix~\ref{sec:AppendixA}):
\begin{equation}
    \expval{ \hat{\alpha}_R}(\tau) \propto 1-\sin \left[ \phi^{(2)}(\tau) \right],
    \label{eq:phi_2}
\end{equation}
where it is assumed that the linear phase $\phi^{(1)}$ is completely averaged out [note that Eq.~\eqref{eq:phi_2} is derived assuming no relaxation]. The measured dependence of the polarization-rotation amplitude $\expval{\hat{\alpha}_R}$ is shown in Fig.~\ref{fig:alphaR_vs_tau}. As predicted above, lengthening of the pulse while keeping its amplitude constant, results in the  $\expval{\hat{\alpha}_R}$ oscillations.
Specifically, for the given experimental conditions, the whole-period amplitude oscillation is observed for the $\SI{70}{\micro\second}$ pulse. For longer pulses, the amplitude of these oscillations deteriorates, which is discussed below.

\begin{figure}
    \centering
    \includegraphics[width=0.95\columnwidth]{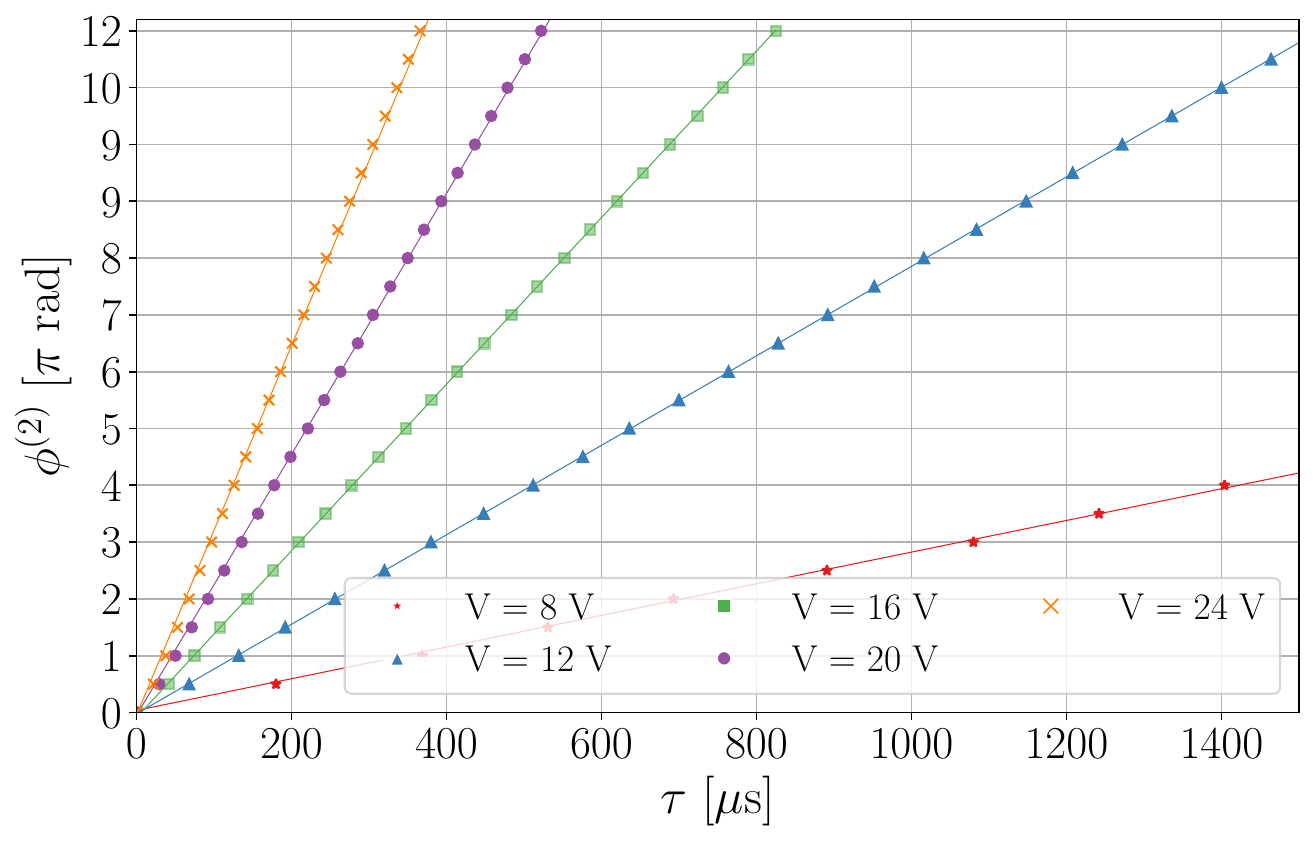}
    \caption{Phase $\phi^{(2)}$ accumulated due to the quadratic Zeeman effect as a function of the pulse duration, for different voltages supplying the H-bridge and hence determining the pulse amplitude. The solid lines are the theoretical predictions and points are measured values. The results show the linear dependence on the pulse length $\tau$ and quadratic dependence on the pulse amplitude $\propto V^2$. }
    \label{fig:phase_accumulation}
\end{figure}

In Fig.~\ref{fig:phase_accumulation}, the measured amplitude of magneto-optical rotation is shown versus the pulse length for a few different magnetic-field-pulse amplitudes (determined by the supply voltage $V$). 
These results confirm the linear dependence of the phase $\phi^{(2)}$ on the pulse length and demonstrate the ability to imprint an arbitrary phase onto atoms. In fact, this enables another way of manipulating the quantum state of the atoms, which will be explored in the context of quantum-state engineering. Simultaneously, one can notice that the steepness of the lines corresponding to different driving voltages rises faster than the voltage itself, eventually confirming the quadratic scaling of the accumulated phase with the magnetic-field amplitude.

As pointed out in Ref.~\cite{Pustelny2006Influence}, in room-temperature atomic vapor, the inhomogeneity of the magnetic field may cause an increase in relaxation. This is because, due to the thermal motion of atoms, the atoms probe an entire volume of the cell, evolving in a varying magnetic field. This random-walk process results in the dephasing of the precesison of spins. A similar behavior is observed in the case of the pulse, if the pulse oscillation frequency is low, the atoms can probe different homogeneities of the field induced with the pulser coils (the effect of the signal-ampltiude deterioration is visible in Fig.~\ref{fig:alphaR_vs_tau}). To avoid this problem, one can use larger coils generating more homogeneous field. Unfortunately, increasing the coils size results in higher impedance, which in turn puts more demands on the parameters of the pulser (particularly, the maximum voltage it needs to generate to inject a specific amount of current into the coils). Alternatively, one can increase the frequency of the oscillating magnetic field, so that the field range, probed by the atom within a single oscillation cycle, is smaller. In fact, this approach allows for the cancellation of the linear Zeeman term. 

\begin{figure}[h!]
    \centering
    \includegraphics[width=0.95\columnwidth]
    {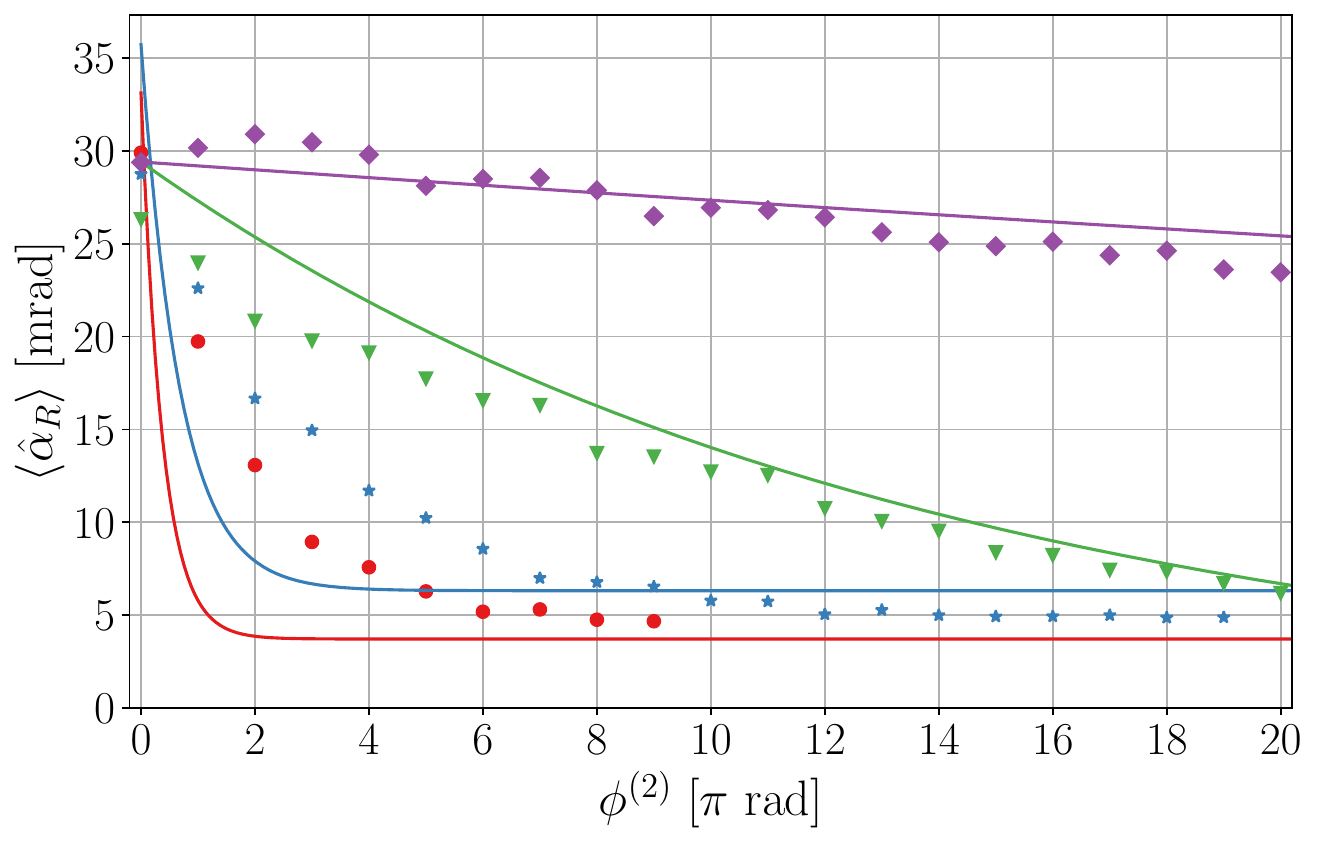}
    \caption{Amplitude $\left\langle \hat{\alpha_R} \right\rangle$ of the polarization rotation versus a number of $\pi$-phase cycle due to quadratic Zeeman effect induced with the pulses of frequencies $\SI{326}{\kilo\hertz}$ (purple), $\SI{200}{\kilo\hertz}$ (green), $\SI{140}{\kilo\hertz}$ (blue), and $\SI{100}{\kilo\hertz}$ (red). Points correspond to the experimental measurements, while the solid lines represent numerical simulations of the effect of the field homogeneity (see main text). Less points shown for $100$ and $\SI{140}{\kilo\hertz}$ originates from low signal-to-noise ratio of the corresponding experimental signals, which disallows us to extract the amplitude.}
    \label{fig:OAO}
\end{figure}

To validate the effect of the oscillation frequency on the polarization dephasing, we conduct a series of simulations and measurements. For the simulations, we perform a series of Monte Carlo simulations in which we consider 10$^4$ particles moving ballistically between the walls of the spherical cell. The velocities of the particles are randomized according to the Maxwell distribution, and the directions of the particles after the wall collision is randomly chosen from the uniform distribution \cite{Zhivun2016}. The magnetic field generated by the coils is calculated by assuming that the Helmholtz coils are built from a wire of negligible width. To simplify the calculation, the evolution of atoms due to the linear Zeeman effect along the $y$-axis is simulated using the optical Bloch equations (not quantum-mechanically with the density matrix) \cite{Colangelo2013}, while the effect of the nonlinear term is approximated by calculating the total accumulated phase from the $y$-component of the field and omitting the effect of the perpendicular components (as transverse components are just second-order corrections). In the measurements, we apply pulses of lengths that allow the accumulation of $\phi^{(2)}=n \pi$ phases, where $n$ is an integer, and the amplitude of magneto-optical rotation $\left\langle \hat{\alpha_R} \right\rangle $ is measured. As shown in Fig.~\ref{fig:OAO}, increasing the frequency of the oscillation leads to a significant reduction of the polarization deterioration. Although there is significant improvement in relaxation when increasing the pulse frequency from 100 to $\SI{326}{\kilo\hertz}$, one can argue that a further frequency increase might reduce the problem even more. However, due to technical limitations this becomes challenging due to the necessity of applying larger supply voltages. This is the reason why most of our measurements are performed at frequencies above $\SI{300}{\kilo\hertz}$ but not higher. 

The results of the simulations (solid lines) and the measurement data (points) are presented in Fig.~\ref{fig:OAO}. Although the simulations do not exactly predict the dependence of the experimental points, especially at lower frequencies, they effectively capture the trends in the data. This indicates that dephasing caused by the linear term of the magnetic field-atom interaction is substantial and contributes to the reduction of the amplitude of the observed signals. The problem is alleviated by using higher frequencies of the oscillating field (see purple data in Fig.~\ref{fig:OAO}).

\section{Conclusion}

In summary, we have developed a novel method to induce the quadratic Zeeman effect, effectively compensating for its linear counterpart. Based on our simulations and experimental data, the system supports the imprinting of the arbitrary phase $\phi^{(2)}$ that is associated with the square of the projection angular momentum operator, $\hat{F}_y^2$, which opens avenues for various quantum-mechanical applications. Specifically, we have demonstrated the ability to imprint the $\phi^{(2)}$ phase up to $12\pi$ between magnetic sublevels of the $F=1$ ground state of $^{87}$Rb atoms [Eq.~\eqref{eq:QuantumMechanicalRotation}]. Importantly, this approach allows one to generate an arbitrarily quantum state within any angular-momentum level (see also an alternative approach developed in Ref.~\cite{Chaudhury2007Quantum}). Traditionally, such a capability required selective coupling among magnetic sublevels and introduced nonlinearity through interactions like the nonlinear Zeeman effect or AC Stark shift, yet in the presence of the dominant linear Zeeman effect. Our method successfully counteracts this effect, thereby enhancing control over the quantum system. This method will find applications in quantum-state and quantum process tomography in system with $F\geq 1$.

It is also noteworthy that the nonlinearity inherent in the quadratic Zeeman effect enables our method to explore various interesting phenomena. For example, we intend to apply this technique for single-axis spin squeezing, which is valuable for reducing quantum noise along one axis of the spin state, thus enhancing quantum metrology measurements. Research into single-axis squeezing is currently taking place in our group.

\begin{acknowledgments}
The authors would like to thank Piotr Put for stimulating discussions at a preliminary stage of the project. This research was financially supported by the National Science Centre of Poland within the Sonata Bis program (grant No. 2019/34/E/ST2/00440).
\end{acknowledgments}

\appto\appendix{\counterwithin{equation}{section}}
\appto\appendix{\counterwithin{figure}{section}}
\appendix

\section{Magneto-optical rotation\label{sec:AppendixA}}

In our experiment, the quantum state of atoms is determined using the linear Faraday effect, as described in detail in Ref.~\cite{marek}. Specifically, at the probing stage, the atoms are subjected to the longitudinal ($z$-oriented) static magnetic field and weak ($<\SI{1}{\micro\watt}$) $y$-polarized probe light, propagating along the $z$-axis. Under such conditions, the light polarization exhibits two oscillating and one non-oscillating components, decaying over time \cite{marek}
\begin{eqnarray}
    \label{eq:FID_signal} 
    \delta \alpha (t; \Delta) &=& \chi e^{-\gamma t} \left[ \expval{\hat{\alpha}_{R}}V_R(\Delta)\sin (2\Omega_L t) \right.\\&\quad+&\left. \expval{\hat{\alpha}_{I}}V_R(\Delta)\cos(2 \Omega_L t) - \expval{\hat{\beta}} V_I(\Delta) \right],\nonumber
\end{eqnarray}
where $\hat{\alpha}_{R(I)}$ and $\hat{\beta}$ are observables, given by the energy-level structure of the atoms. It can be shown that for the $F=1$ ground state, the observables can be written as \cite{marek}
\begin{eqnarray}
    \hat{\alpha}_R & = & A (\dyads{1,1}{1,-1}{z} + \dyads{1,-1}{1,1}{z}),\nonumber\\
    \hat{\alpha}_I & = & i A (\dyads{1,-1}{1,1}{z} - \dyads{1,1}{1,-1}{z}),\label{eq:observable_z_basis}\\
    \hat{\beta} & = & B \left(\dyads{1,1}{1,1}{z} - \dyads{1,-1}{1,-1}{z} \right),\nonumber
    \label{eq:Observables}
\end{eqnarray}
where $A$ and $B$ are constants given by the total angular momentum $F'$ of the excited level. Since the discussion below only concerns the spin-1 atoms, hereafter we omit the total angular momentum of the state in further discussions, i.e., $\ket{1,m_{F}} \rightarrow \ket{m_F}$.
\begin{figure}
    \centering
    \includegraphics[width = 0.9\columnwidth]{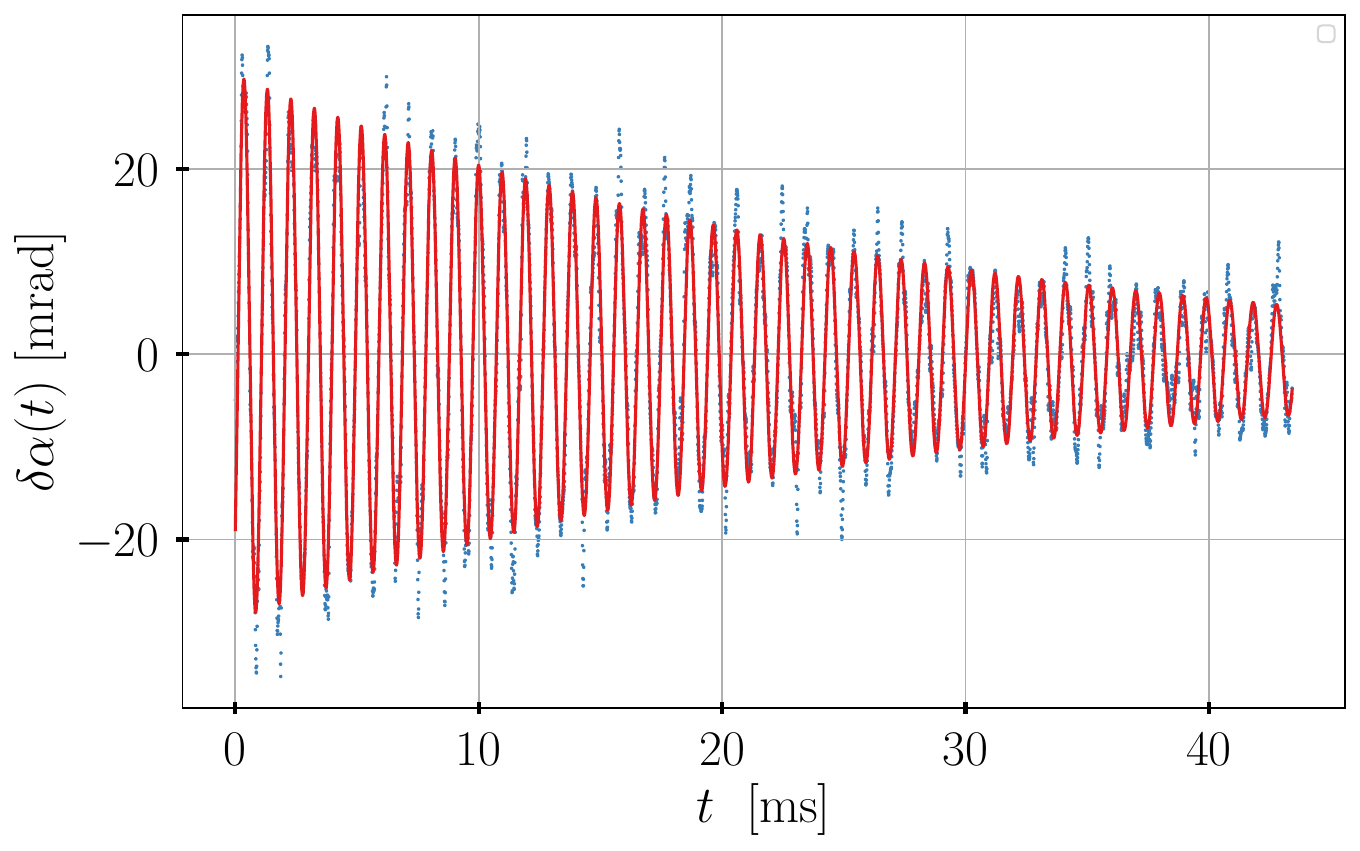}
    \caption{Example of experimentally measured signal (blue points) with the fit (red line) according to the Eq.~\eqref{eq:FID_signal}. The presented signal was measured for the state \eqref{eq:initial_state} with additional $\pi/4$ rotation around $x$ axis [see Eq.~\eqref{eq:state_z_basis}].}
    \label{fig:FID}
\end{figure}

For the evaluation of our method, we utilized the $\ket{1}_{x}$ state. This state can be prepared with circularly polarized light propagating along the $x$-axis. For the quantization axis along $z$, the state is given by
\begin{equation}
\begin{split}
    \ket{\psi(0)} =& \ket{1}_{x} = \hat{O}_{x\rightarrow z}\ket{1}_{z} \\  =&\dfrac{1}{2}\left( \ket{1}_{z} + \sqrt{2}\ket{0}_{z} + \ket{-1}_{z} \right),
\end{split}
\label{eq:initial_state}
\end{equation}
where $\hat{O}_{x \rightarrow z}$ is the operator rotating the system quantization axis from $x$ to $z$. 

As the energy-level shift can be described by Eq.~\eqref{eq:LevelSplitting}, one can write the evolution operator associated with the magnetic-field pulse in the $\ket{m_F}$ basis
\begin{eqnarray}
    \left[U_{\tau}(T)\right]_{m_F} & = & \exp{-i\int_{0}^{T} \Omega_{L}^{(1)}(t') m_{F} + \Omega_{L}^{(2)}(t') m_{F}^{2} \dd{t'}}\nonumber\\
    &\approx& \exp{-i m_{F}^{2} \int_{0}^{T} \Omega_{L}^{(2)}(t') \dd{t'} }\label{eq:time_evolution} \\
    &\approx& \exp{-i \phi^{(2)}(\tau) m_{F}^{2}}.\nonumber
    \label{eq:PulseOperator}
\end{eqnarray}
Here, the second line assumes that the operator part associated with the linear Zeeman effect averages out, which is true for sufficiently long $T$ [see Fig.~\ref{fig:Zeeman_effect}]. A long $T$ also justifies the third line and the replacement of the integral of $\Omega_L^{(2)}$ over time with $\phi^{(2)}(\tau)$ [see Eq.~\eqref{eq:phase_def}]. 
The time $T$ can be roughly estimated using the decay rate $\Gamma$ of the $RLC$ circuit. For example, for $T = \tau + 10/\Gamma$ the amplitude of the remaining oscillation is damped to a level of $5\times 10^{-5}$ of the maximum pulse amplitude.

It is important to note that, as the eigenenergies of the system are described using Eq.~\eqref{eq:LevelSplitting}, the corresponding evolution operator is diagonal and fully described by Eq.~\eqref{eq:time_evolution}. This allows us to write an explicit form of the time evolution operator 
\begin{equation}
    \hat{U}_{\tau}(T) = e^{-i \phi^{(2)}} \left( \dyads{1}{1}{y}+ \dyads{-1}{-1}{y} \right) + \dyads{0}{0}{y},
    \label{eq:QuantumMechanicalRotation}
\end{equation}
where the subscript indicates that the state is given with the quantization axis oriented along the direction of the oscillating magnetic field ($y$-axis). Note that from now on, $\phi^{(2)}\equiv\phi^{(2)}(\tau)$.

The operator \eqref{eq:QuantumMechanicalRotation} is given in the eigenbasis oriented along the $y$-axis, which is not the basis in which the observables~\eqref{eq:Observables} are defined. Therefore, to maintain the consistency of the calculations, we rotate the operator toward the $z$-axis
\begin{widetext}
\begin{eqnarray}
    \hat{U}_{\tau}(T) &=& \hat{O}_{y\rightarrow z}\left[ e^{-i \phi^{(2)}} \left( \dyads{1}{1}{y}+ \dyads{-1}{-1}{y} \right) + \dyads{0}{0}{y} \right]\hat{O}_{y\rightarrow z}^{\dagger}\\
    &=&\dfrac{1+e^{-i \phi^{(2)}}}{2}\left( \dyads{1}{1}{z} + \dyads{-1}{-1}{z}\right)
    + \dfrac{e^{-i \phi^{(2)}}}{2}\dyads{0}{0}{z} + \dfrac{e^{-i \phi^{(2)}}-1}{2} \left(\dyads{1}{-1}{z} + \dyads{-1}{1}{z} \right).\nonumber
\end{eqnarray}
\end{widetext}

This form of the operator may be used in our calculations.

The last stage of our state manipulation prior to probing is the rotation of the state by $\pi/4$ around the $x$-axis. In turn, the final state $\ket{\psi_{\tau}(T)}$ of the atoms is given by
\begin{eqnarray}
    \ket{\psi_{\tau}(T)} &=& \hat{R}^{(x)}\left(\dfrac{\pi}{4}\right) \hat{U}_{\tau}(T) \ket{\psi(0)} \label{eq:state_z_basis}\\
    &=& \dfrac{1}{2\sqrt{2}}\left(1+ i e^{-i \phi^{(2)}}\right) \left( \ket{1}_{z} + \ket{-1}_{z}\right)\nonumber\\ &&\qquad + \dfrac{1}{2}\left(-i + e^{-i \phi^{(2)}}\right) \ket{0}_{z}\nonumber.
\end{eqnarray}

Using this state and utilizing Eqs.~\eqref{eq:state_z_basis} and \eqref{eq:observable_z_basis}, one can calculate the expectation value of the observables arising from Eq.~\eqref{eq:FID_signal}
\begin{eqnarray}
    \expval{\hat{\alpha}_R} &=& \ev**{\hat{\alpha}_R}{\psi_{\tau}(T)} \propto \left\lbrace 1-\sin \left( \phi^{(2)} \right) \right\rbrace,\nonumber\\
    \expval{\hat{\alpha}_i} &=& 0,\\
    \expval{\hat{\beta}} &=& 0.\nonumber
\end{eqnarray}
We use the first of these equations to analyze the build-up of the quadratic phase evolution $\phi^{(2)}$ and thus characterize our experimental method.

\bibliography{apssamp.bib}

\end{document}